\documentstyle[12pt]{article}

\newcommand{\bfeps}{\mbox {\boldmath $\epsilon$}}
\newcommand{\bfsigma}{\mbox {\boldmath $\sigma$}}
\newcommand{\dsdo}{\frac {d\sigma}{d\Omega}}

\newcommand{\aofo}{$a^0_0$$\,\leftrightarrow\,$$f_0$}
\newcommand{\mix}{$a^0_0\,$$-$$f_0$}
\newcommand{\pieta}{$\pi^0$$\,\leftrightarrow\,$$\eta$}
\newcommand{\aofomix}{\mbox {\boldmath $a_0\,$-$f_0$}}
\newcommand{\pndao}{\mbox {\boldmath $pn\to d a_0$}}

\date{}

\begin{document}

\centerline{\Large \bf On The Possibility of Observation of
  \aofomix\ Mixing}
\centerline{\Large \bf in the \pndao\  Reaction\footnote{
 This is the modified version of the paper published in JETP Lett. {\bf 72}
 (2000) 410.}}

\vspace{5mm}

\centerline{\bf A.E.Kudryavtsev, V.E.Tarasov}

\vspace{3mm}

\centerline{\it Institut of Theoretical and Experimental Physics}
\centerline{\it B.Cheremushkinskaya 25, 117259 Moscow, Russia}
\centerline{\it e-mail: kudryavtsev@vxitep.itep.ru,
                        tarasov@vxitep.itep.ru}
\vspace{3mm}
\begin{abstract}
It is shown that, if isospin is not conserved in $a^0_0$- and
$f_0$-meson mixing, forward-backward asymmetry arises in the reaction
$pn\to da^0_0$. This effect increases near the reaction threshold.
The asymmetry is estimated within the framework of a model in which
the \mix\ mixing is due to the virtual \pieta\ transition and the
difference in masses of the charged and neutral kaons in decay channels.
The angular asymmetry near the threshold of the $pn\to da^0_0$ reaction
was found to be large, of the order of $8\div 15$ \%. \\ \\
{\bf PACS}: 13.60.Le, 13.75.-n, 14.40.Cs 
\end{abstract}

The origin of the lightest, virtually mass-degenerate, scalar mesons
$a_0$ (980) ($I^G J^{PC}=1^- 0^{++}$) and $f_0$ (980) ($0^+ 0^{++}$)
is one of the most important problems of hadron physics. Different
assumptions exist about the structure of these mesons, from the
standart $q\bar q$ states~\cite{Torn} and their modifications
(see, e.g.,~\cite{Ani} and reference therein) to the 4-quark
configurations~\cite{Acha} and the lightest scalar mesons as "minions"
in the Gribov confinement model~\cite{Grib}. The problem of the
structure of $a_0$ and $f_0$ mesons is closely related to the problem
of $a_0\,$-$f_0$ mixing. The dynamical mechanism of this mixing was
suggested around 20 years ago in Ref.~\cite{Acha1}.
The first evidence of this phenomena came from the recent data of
the WA102 collaboration at CERN~\cite{Barb}. These data on the central
production of $a_0$ and $f_0$ in the reaction $pp\to p_s M p_f$ were
interpreted in Ref.~\cite{Close} as the evidence of the $a_0\,$-$f_0$
mixing.  

If the $a_0$ and $f_0$ mesons have close structures, then the mixing
with violation of isospin conservation could be large. Along with the
direct \aofo\ transition due to isospin violation in the quark sector,
these mesons can mix due to isospin-violating interaction in the decay
channels. Different mixing mechanisms are illustrated in Fig.~1.
Note that the vertex of the direct \mix\ interaction in Fig.~1a depends
on the quark content of scalar mesons and should be extracted from the
experiment. At the same time, the mixing due to the decay processes
presented in Fig.~1b and 1c can be estimated rather reliably.

It is convenient to examine \mix\ mixing in the reaction of production
of a neutral $a_0$ meson: 
\begin{equation} p\,n \to d\, a^0_0\, . \label{1}\end{equation}
Note that the forward-backward asymmetry in reaction~(\ref{1}) is
absent if the isospin is conserved~\cite{Miller}. As will be shown
below, observation
of this asymmetry would testify to the presence of isospin-violating
\mix\ mixing, and the asymmetry effect should be stronger near the
threshold of the reaction~(\ref{1}).

If isospin is conserved, an isovector $a_0$ meson can be produced near
the threshold of the reaction~(\ref{1}) only in the $p$ wave with
respect to the deutron. At the same time, if an isoscalar $f_0$ meson
is produced in the reaction
\begin{equation} p\,n \to d\, f_0\, , \label{2}\end{equation}
the final orbital angular momentum $L$ of the $df_0$ system may be zero.
This conclusion follows from the isospin ($I$), parity ($P$), and
angular momentum ($J$) conservation laws. The possible quantum numbers
for reactions~(\ref{1}) and~(\ref{2}) yielding final systems with the
smallest orbital angular momenta ($p$ and $s$ waves for the $d\,a^0_0$
and $d\,f_0$ systems, respectively) are listed in the table. The total
spin of the system is denoted by $S$. The quantum numbers presented in
the table are consistent with the requirement for antisymmetry of the
system with respect to the initial fermions.
\begin{center}
\begin{tabular}{|c|c|c|c|c|}
\multicolumn{5}{c}{{\bf Table}} \\
\hline
& \multicolumn{2}{|c|}{$pn\to d\,a_0$}
& \multicolumn{2}{|c|}{$pn\to d\,f_0$} \\
\hline
I & 1 & 1 & 0 & 0 \\ S & 1 & 1 & 1 & 1 \\
L & 1, $\phantom{.}$3 & 1 & 0, $\phantom{.}$2 & 0 \\
P & -1 & -1 & 1 & 1 \\
\hline
\end{tabular} \end{center}
Thus, if isospin is conserved, reactions~(\ref{1}) and~(\ref{2}) should
have different energy and angular dependences. In particular, for the
near-threshold production of stable mesons, one has
\begin{equation}
\sigma(pn\to d a^0_0) \sim Q^{3/2}\, , \phantom{xxx}
\sigma(pn\to d f_0) \sim Q^{1/2}
\phantom{xx} (Q=\sqrt{s} -m_d -\bar m)\, ,
\label{3}\end{equation}
where $Q$ is the energy release in the respective reaction; $\sqrt{s}$
is the total CM energy; and $m_d$ and $\bar m$ are the deutron and
meson masses, respectively.

Let us assume that the $f_0\to a^0_0\,$ transition can proceed without
isospin conservation. Then, the $a_0$ meson in reaction~(\ref{1}) can
be produced in the $s$ wave with respect to the final deutron. As is
seen from the table, the initial spin state of nucleons in both
reactions $pn\to d a^0_0\,$ and $pn\to d f_0\to d a^0_0$ is the same
($S=1$). Therefore, the $p$-wave amplitude of the main
process~(\ref{1}) interferes with the $s$-wave amplitude of the
isospin-violating process $pn\to d f_0\to d a^0_0\,$. Due to this
interference, an asymmetry arises in the forward-backward escape of
the $a_0$ meson in reaction~(\ref{1}). In this case, the process with
isospin conservation is energetically suppressed in the range of low
$Q$ values, as follows from Eqs.~(\ref{3}). For this reason, the
angular asymmetry in the $a_0$-meson production may be large near the
threshold.\footnote{
Note that, for the $pn\to d\pi^0$ reaction discussed in~\cite{Nisk},
the angular asymmetry of the near-threshold cross section is suppressed.
This is due to the fact that the process with isospin conservation
yields the $d\pi^0$ system in the $s$ wave.} 

The asymmetry $A\,$ for reaction~(\ref{1}) is defined as
\begin{equation}
A=\frac{\sigma_+ -\sigma_-}{\sigma_+ +\sigma_-}\, ,
\phantom{xx}
\sigma_{\pm} = \dsdo (z=\pm 1)\, ,
\phantom{xx}
z=\cos \theta\, ,
\label{4}\end{equation}
where $\theta$ is the polar CM angle of $a^0_0$-meson escape, with
the polar axis coinciding with the initial beam.

For the numerical estimations of asymmetry $A\,$, we first consider
the $\Lambda_{af}$ vertex determined by the diagram in Fig.~1b for the
$a^0_0\to f_0\,$ transition. The $\lambda_{\pi\eta}$ vertex
corresponding to the \pieta\ transition in this diagram is known from
the theoretical analysis of the $\eta\to 3\pi^0$ reaction~\cite{Coon}
(see also~\cite{Nisk}). We take $\lambda_{\pi\eta} \simeq -5000$ MeV$^2$
as an estimate, which is the average of the theoretical values given
in~\cite{Nisk,Coon}. Direct calculation of the diagram in Fig.~1b
yields the following result for the contribution $\Lambda_{\pi\eta}$
of the process indicated in Fig.~1b to $\,\Lambda_{af}\,$:
\begin{equation}
\Lambda_{\pi\eta} =\frac{\lambda_{\pi\eta}\, g_{a\pi\eta}\,
g_{f\pi^0\pi^0}}
{16\pi^2 {\bar m}^2} \left( \frac{{\bar m}^2}{m^2_{\eta}}
\ln \frac{{\bar m}^2 - m^2_{\eta}}{{\bar m}^2}
-\ln \frac{{\bar m}^2 - m^2_{\eta}}{m^2_{\eta}} +i\pi \right)
\approx (118 - 186\, i)\, {\rm MeV}^2
\label{5}\end{equation}
Here, $m_{\eta}$ is the $\eta$ mass, $\bar m =$980~MeV/$c^2$ is the
mass of the $a_0$ and $f_0$ mesons, and $\,g_{a\pi\eta}$ and
$g_{f\pi^0\pi^0}$ are the vertices of the $a_0\to \pi\eta$ and
$f_0\to 2\pi^0$ decays. Estimate~(\ref{5}) was obtained with zero pion
mass $m_{\pi}=0$ and under the assumption that the width of the
$a_0$ and $f_0$ mesons with nominal mass $\bar m$ are 
$\Gamma (\bar m) \equiv \Gamma_0 =50$~MeV/$c^2$ and determined only by
the decays through the $\pi\eta$ and $\pi\pi$ channels, respectively.
Then, $g^2_{a\pi\eta}=8\pi{\bar m}^2 \Gamma_0/q_{\pi\eta}\,$,
$\,g^2_{f\pi\pi}=8\pi{\bar m}^2 \Gamma_0/q_{\pi\pi}\,$ and
$g_{f\pi^0\pi^0}=g_{f\pi\pi}/\sqrt{3}$, where $q_{\pi\eta}\,$ and
$q_{\pi\pi}\,$ are the relative momenta in the $\pi\eta$ and $\pi\pi$
systems. For the \mix\ mixing angle, Eq.~(\ref{5}) gives the estimate
$\sin\theta_{af}\simeq |\Lambda_{\pi\eta}/{\bar m}\Gamma_0 |\simeq 0.0045$.

The mechanism of external mixing due to the $K\bar K$ decay channel,
discussed in Ref.~\cite{Acha1} and in recent paper~\cite{Kerb}, shows
that the kinematic isospin violation due to the difference in masses of
the $K^{\pm}$ and $K^0$ mesons is large and considerably stronger than
that due to the $\pi$$-$$\eta$ mixing. At the same time, the strong
isospin violation is concentrated in a narrow range of the $a_0$-meson
masses near the thresholds of the decays through the $K\bar K$ channels.

The vertex $\Lambda_{K\bar K}$ corresponding to the \aofo\ transition
(Fig.~1c) has the form
\begin{equation}
\Lambda_{K\bar K}(\bar m) =\frac{g_{a K\bar K}\, g_{f K\bar K}}{32\pi}
 i \left( \sqrt{\frac{{\bar m}-2m_{K^+}}{m_K} +i0}\,
 - \sqrt{\frac{{\bar m}-2m_{K^0}}{m_K} +i0}\,\right),
\label{6}\end{equation}
where $m_{K^+}=493.7$~MeV/$c^2$ and $m_{K^0}=497.7$~MeV/$c^2$ are the
masses of, respectively, charged and neutral kaons~\cite{PDG};
$m_K=(m_{K^+} + m_{K^0})/2$; and $g_{a K\bar K}$ and $g_{f K\bar K}$
are the vertices of the $a_0\to K\bar K$ and $f_0\to K\bar K$ decays,
respectively. In addition, $g^2_{a K\bar K}=2g^2_{a K^+K^-}\,$,
$\,g^2_{f K\bar K}=2g^2_{f K^+K^-}\,$,
$\,g_{a K^0\bar K^0}=-g_{a K^+K^-}\,$, and
$\,g_{f K^0\bar K^0}=g_{f K^+K^-}$. To numerically estimate
Eq.~(\ref{6}), we set $g_{aK\bar K}=g_{a\pi\eta}\,$, in agreement with
the experimental restriction
$g^2_{aK\bar K} /g^2_{a\pi\eta}=0.91\pm 0.11$~\cite{Teige}.
We also assume that $\,g^2_{f K\bar K}=g^2_{a K\bar K}\,$ and
$\,g_{a K^+ K^-} = -g_{f K^+K^-}\,$ (the relative signs of the $a_0$-
and $f_0$-meson decay vertices are consistent with the predictions of
SU(3) symmetry).\footnote{
It is assumed that the $a_0$ isotriplet and $f_0$ isosinglet belong to
the same scalar meson octet.}
The function $\Lambda_{K\bar K}(m)$ is shown in Fig.~2. This function
sharply changes near the thresholds of the $K^+K^-$ and $K^0\bar K^0$
systems within the mass intervals comparable to the 
$2m_{K^0}-2m_{K^+}\approx 8$~MeV/$c^2\,$ difference, which is
considerably smaller than the width of the $a_0$ and $f_0$ mesons.
It is seen that $|\Lambda_{K\bar K}|\gg |\Lambda_{\pi\eta}|$
near the $K\bar K$ thresholds.

Let us assume that the signal from the $a^0_0\,$-meson production is
detected by identifying the $\pi^0\eta$ final state. In this case,
the use of perturbation theory for the isospin-violating interaction
gives three diagrams dominating the cross section for the process.
They are shown in Fig.~3. The $M_1$ diagram corresponds to the process
with isospin conservation. The $M_2$ and $M_3$ diagrams are first-order
corrections in the isospin-violating interaction to the diagram $M_1$.
The $M_2$ diagram is irrelevant the $a_0\,$-meson production but does
contribute to the $s$$-$$p$ interference under discussion.
This contribution corresponds to the interference of diagrams $M_2$ and
$M_1$ through the \mix\ mixing mechanism (Fig.~1b). The respective
amplitudes $M_{1,2,3}\,$ are
\begin{equation}
M_1=M_a\, G_a\, g_{a\pi\eta}\, ,
\phantom{xx}
M_2=M_f\, G_f\,
\frac{g_{f\pi^0\pi^0}\,\lambda_{\pi\eta}}{m^2_{\eta}-m^2_{\pi}}\, ,
\phantom{xx}
M_3=M_f\, G_f\, \Lambda_{af}\, G_a\, g_{a\pi\eta}\, ,
\label{7}\end{equation}
where
$\Lambda_{af}=\Lambda_{dir} +\Lambda_{\pi\eta} +\Lambda_{K\bar K}$
is the vertex corresponding to the \aofo\ transition. The vertex
$\Lambda_{dir}$ of direct interaction (Fig.~1a) was not taken into
account; i.e., $\Lambda_{af}=\Lambda_{\pi\eta} + \Lambda_{K\bar K}$.
In Eqs.~(\ref{7}), $G_a$ and $G_f$ are the propagators of the
$a_0$ and $f_0$ mesons, respectively:
\begin{equation}
G_a=G_f=\frac{1}{2\bar m}\cdot \frac{1}{m-\bar m +i\Gamma(m)/2}\, ,
\phantom{xx}
\Gamma (m) = \Gamma _0 + \frac{g^2_{aK\bar K}}{8\pi {\bar m}^2}\,
\sqrt{m_K (m-2m_K)+i0}\, ,
\label{8}\end{equation}
where the width $\Gamma(m)$ takes into account that the resonances may
decay through the $K\bar K$ channel~\cite{Flat} and $\,m$ is the mass
of the $\pi\eta$ system. In Eqs.~(\ref{7}), $M_a$ and $M_f$ denote the
amplitudes of production of the $a_0\,$ and $f_0\,$ mesons in
reactions~(\ref{1}) and~(\ref{2}), respectively. These amplitudes can
be estimated from the diagrams of the impulse approximation (Fig.~4).
With allowance made only for the $s$-wave component of the deuteron
wave function (WF), the expressions for $M_a$ and $M_f$ antisymmetrized
over the initial nucleons have the form
\begin{equation}
M_a =g_{aNN}\,\sqrt{m_N}\,\left( u(q_1)-u(q_2)\right)\, X\, ,
\phantom{xx}
M_f =g_{fNN}\,\sqrt{m_N}\,\left( u(q_1)+u(q_2)\right)\, X\, .
\label{9}\end{equation}
Here, $g_{aNN}$ and $g_{fNN}$ are the vertices for the $a_0\,$- and
$f_0\,$-meson coupling to the nucleon; $m_N$ is the nucleon mass;
$u(q)$ is the deuteron WF; $q_1$($q_2$) is the relative momentum in the
deuteron vertex corresponding to the emission of an $a^0_0$ or $f_0$
meson by the initial proton (neutron); and
$X =\varphi^T_p\sigma_2\bfeps\cdot\bfsigma\varphi_n$ is the spin factor,
where $\varphi_p$ and $\varphi_n$ are the proton and neutron spinors,
respectively, and $\bfeps$ is the deuteron polarization vector.
Let ${\bf p}$ and ${\bf k}$ be the CM 3-momenta of the initial proton
and the final $\pi\eta$ system, respectively. For $m\simeq \bar m$,
one has $k \ll p$ and 
$q^2_{1,2}\approx p^2 \pm (E_N/m_N)({\bf p}\cdot {\bf k})$
near the threshold $Q=\sqrt{s}-m_d-\bar m\simeq 0$
of the reaction~(\ref{1}), where $E_N\approx m_N+\bar m /2$ is the total
nucleon CM energy. Then, Eqs.~(\ref{9}) take the form
\begin{equation}
M_a =2\,g_{aNN}\,\sqrt{m_N}\,
\frac{du(p)}{dp^2}\,\frac{E_N}{m_N}\,({\bf p}\cdot {\bf k})\cdot X\, ,
\phantom{xx}
M_f =2\,g_{fNN}\,\sqrt{m_N}\, u(p)\cdot X\, .
\label{10}\end{equation}
It follows from these expressions that at small $k$ values the
amplitude $M_a$ of $a_0$-meson production is much smaller than the
amplitude $M_f$ of $f_0$-meson production. Note that the relative
contributions of the $p$- and $s$-wave amplitudes $M_a$ and $M_f$ to
the cross section for reaction~(\ref{1}) depend on the width of the
$a_0$ and $f_0$ mesons and on the restrictions on the mass range of the
final $\pi\eta$ system. In what follows, we set $g_{aNN}\equiv g_{fNN}$
in Eq.~(\ref{10}). For the deuteron WF with momenta $p\sim 1$~GeV/c, we
take $u(p)\sim p^{-n}$. The Hulthen WF corresponds to $n=4$; i.e.,
$du(p)/dp^2 = -2\,p^{-2}u(p)$. This approximation for the deuteron WF
and the impulse approximation for the amplitudes $M_a$ and $M_f$ cannot
give reliable estimates for the absolute values of the cross sections
and are used only for estimating the asymmetry $A$. The differential
cross section for the reaction $pn\to d\pi^0\eta$ can be written as
\begin{equation}
\dsdo =N \int\limits_{m_{min}}^{m_{max}} |M_1+M_2+M_3|^2\, k\,dm\, .
\label{11}\end{equation}
The $m$ dependence of the amplitudes $M_{1,2,3}$~(\ref{7}) is taken into
account for $\Lambda_{K\bar K}$~(Eq.(\ref{6})); $G_a\,$, $G_f$ and
$\Gamma (m)$~(Eqs.(\ref{8})); and for the relative momentum
$k\,$=$\,\sqrt{2\mu (Q+\bar m -m)}$, where
($\mu = m_d \bar m/(m_d + \bar m)$). To a common factor, the amplitudes
$M_a$ and $M_f$~(\ref{10}) for the unpolarized particles can be written
as $M_a= -2(E_N/m_N) (k/p)\,z$ and $M_f=1$. The normalization constant
$N$ in Eq.~(\ref{11}) contains weakly varying factors, and its magnitude
is of no interest to us. The quantities $p$ and $E_N$ are calculated at
the threshold; i.e., $E_N=m_N+\bar m/2$ and
$p=\sqrt{\bar m (m_N+\bar m/4)}\simeq 1$ GeV/c.

The width of the mass interval $(m_{min}, m_{max})$ of the $\pi^0\eta$
system (more precisely, its lower limit $m_{min}$), in which the
$a_0$ mesom is detected, is an important factor for estimating
asymmetry $A$~(\ref{4}). When the mass of $\pi^0\eta$ system decreases
(below the nominal mass $\bar m$), the momentum $k$ increases, resulting
in both the enhancement in the $p$-wave amplitude $M_a\,$, as compared
to the $s$-wave amplitude $M_f\,$, and the dependence of the asymmetry
$A$ on $m_{min}$. In what follows, we specify integral~(\ref{11})
between the limits $m_{max}=Q+\bar m$ (kinematic boundary) and
$m_{min}=\bar m -C (\Gamma_0\,/2)$, where $C$ is a variable parameter.
The calculated (by Eqs.~(\ref{4}) and~(\ref{11})) asymmetry $A$ of the
$\pi^0\eta$-system production is shown in Fig.~5 as a function of energy
release $Q$~(\ref{3}). The calculations were carried out for two lower
limits (corresponding to $C=1$ and $C=2$) on the mass of the $\pi^0\eta$
system. A decrease in the effect with increasing parameter $C$ is due
to the increase in the role of the main (isospin-concerving) process
of $a_0\,$-meson production in the $p$ wave. Note that the contribution
of the $p$-wave amplitude to the cross section dominates over the
$s$-wave contribution in both variants if the $a_0\,$-meson width is
taken into account. For this reason, the relative contribution of the
$s$$-$$p$ interference and, hence, the asymmetry, decrease upon
enhancing the $p$ wave. The dashed (dotted) curves in Fig.~5 correspond
to the asymmetry calculations taking account of only one mixing
mechanism given by the diagram in Fig.~1b (Fig.~1c), i.e., for
$\Lambda_{af}$=$\Lambda_{\pi\eta}$
($\Lambda_{af}$=$\Lambda_{K\bar K}$). The diagram $M_2$ (Fig.~2) is
automatically ignored in calculating the dotted curves.

As is seen from our calculations in Fig.~5, the asymmetry in the
nearthreshold $a_0\,$-meson production is rather large
(about $8\div 15$ \%), which enables one to believe that it can be
experimentally observed. Note that nonresonance production of the
$\pi^0\eta$ system in the reaction $pn\to d\pi^0\eta$ looks like
a background to the reaction $pn\to d\,a^0_0\to d\pi^0\eta$.
Fortunately this background gives no contribution to the discussed
asymmetry $A$, see, e.g. Ref.~\cite{Miller} for more details.

Note in conclusion that our estimates of the asymmetry may be improved.
This is primarily true for the calculation of the amplitudes
$M_a$ and $M_f\,$ of the $a_0\,$- and $f_0\,$-meson production.
An approach with the inclusion of diagrams describing intermediate
rescattering processes (see, e.g.,~\cite{Gri}) seems to be more
reliable than the inclusion of the pole diagrams (Fig.~4) that
depend on the behavior of the deuteron WF at high momenta
$\sim 1$~GeV/c, where the WF is poorly known.

There is also a possibility to test the \mix\ mixing in the reaction
$dd\to a^0_0\,{^4}$He, which is forbidden by isospin conservation. 
Note that the analogous forbidden reaction $dd\to \pi^0\,{^4}$He
was not observed. However, it could proceed through $\eta$- and
$\eta'$-production reactions due to $\pi$-$\eta$- and
$\pi$-$\eta'$-mixing mechanisms~\cite{Coon1}.
We expect that the $a_0$-production process in the nearthreshold region
has some privelegies in comparison to the $\pi^0$ production.
Note that the final $a^0_0$ meson may be produced in $s$ wave
with respect to ${^4}$He  from the initial $s$-wave $dd$ state.
That is why the formation of ${^4}$He  wave function should not be 
accompanied by the rearangement of the orbital motion of nucleons.
In the case of $\pi^0$ production the situation is opposite:
$\pi^0$ meson is produced in $s$ wave, initial $dd$ system is in
$p$ wave. That is why the formation of ${^4}$He wave function from
initial $dd$ state is accompanied by rearangement of the orbital motion
of nucleons. Thus, the process of $\pi^0$ production is dynamically
suppressed versus $a_0$ case.

We are grateful to all participants of the Workshop on the
ANKE-Spectrometer Program held at the ITEP (July 2000) and particularly
to K.G.Boreskov and V.P.Chernyshev for helpful discussions and interest
in this study. We also thank N.N.~Achasov for useful discussion and
information about his recent papers, related to this subject.


\newpage
\centerline{\large\bf Figure captions}
\vspace{5mm}

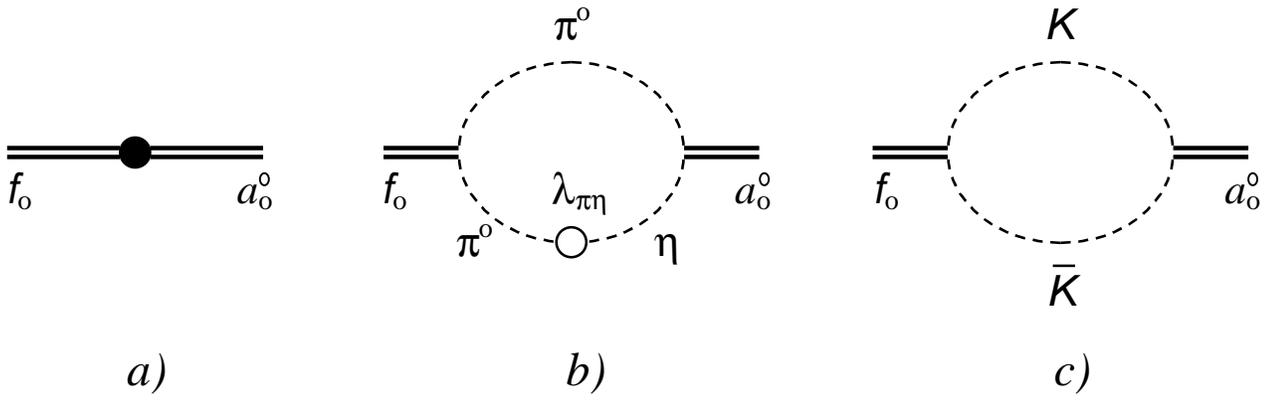
{\bf Fig.~1.} Different types of interactions resulting in the \mix\
     mixing: (a) direct (or contact), (b) due to the virtual \pieta\
     transition, and (c) due to the mass difference between
     $K^{\pm}$ and $K^0$ mesons.
\vspace{3mm}

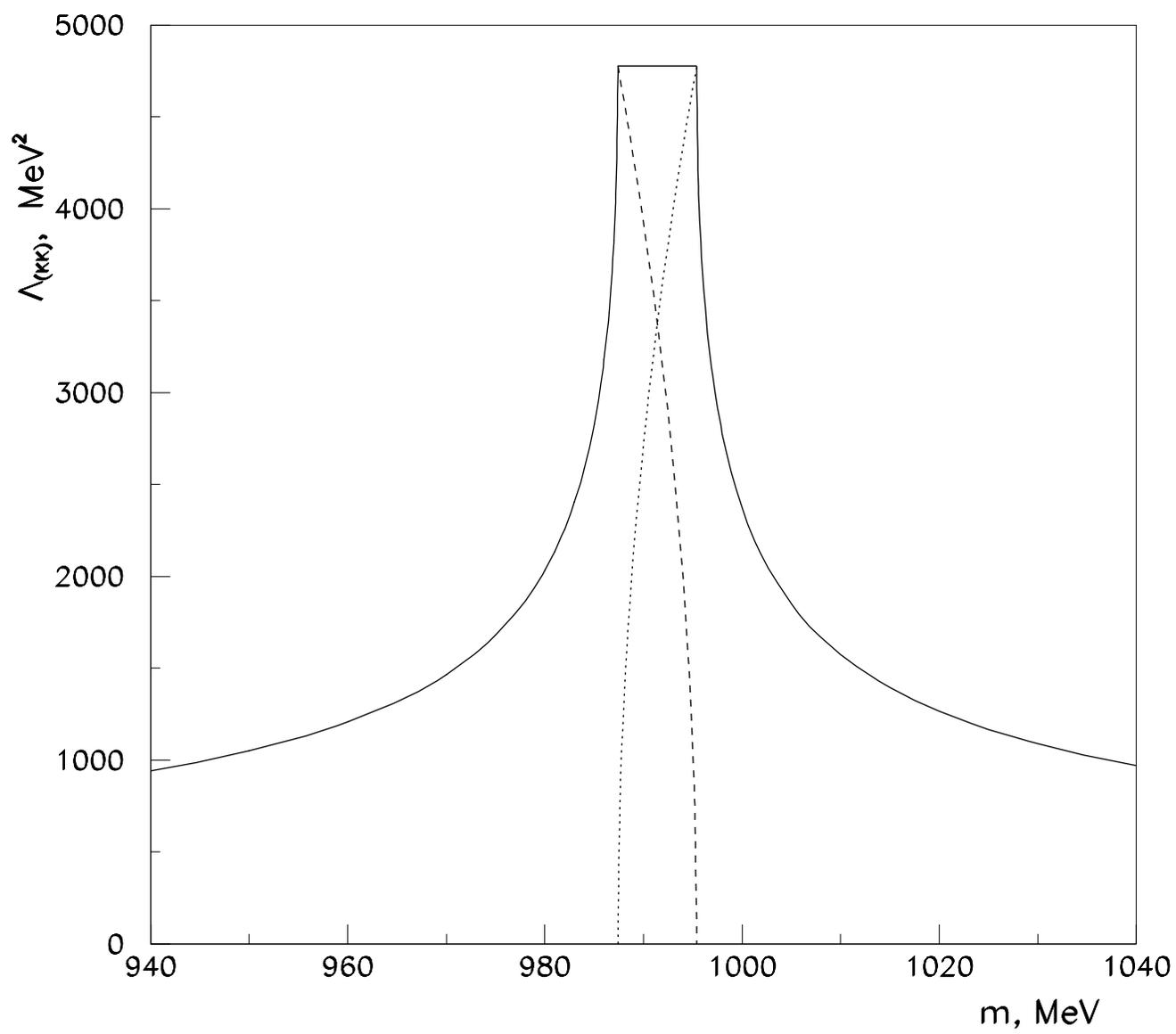
{\bf Fig.~2.} Vertex function $\Lambda_{K\bar K}$~(\ref{6}) vs.
     $a_0\,$-meson mass $m$. The solid, dashed and dotted curves
     correspond to its absolute value, real part, and imaginary part,
     respectively. The dashed (dotted) curve coincides with the solid
     curve for $m<2m_{K^+}$ ($m>2m_{K^0}$).
\vspace{3mm}

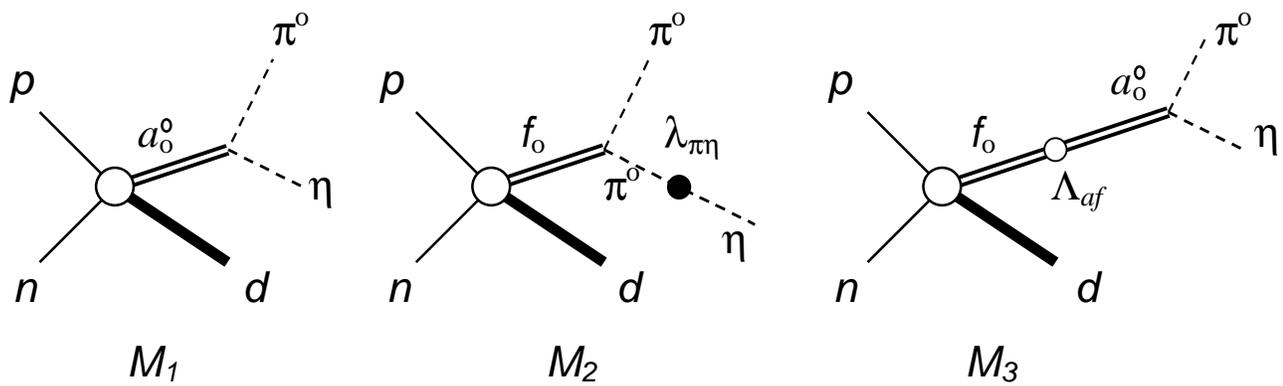
{\bf Fig.~3.} Zero- and first-order isospin-violating diagrams of the
     $pn\to d\pi^0\eta$ process.
\vspace{3mm}

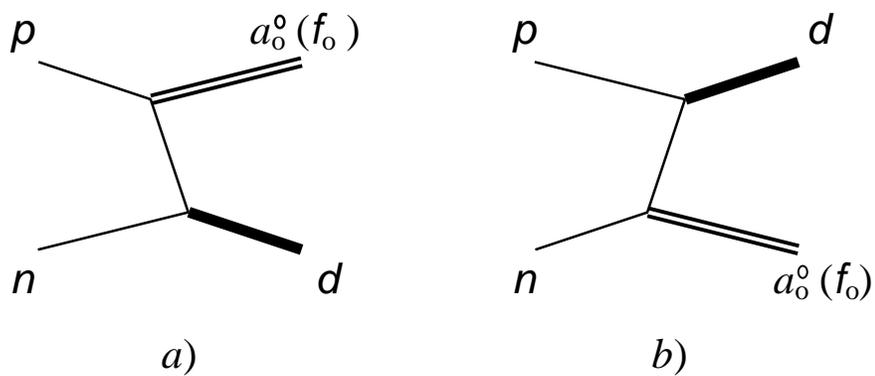
{\bf Fig.~4.} Diagrams for the $a_0$($f_0$)-meson production in impulse
     approximation.
\vspace{3mm}

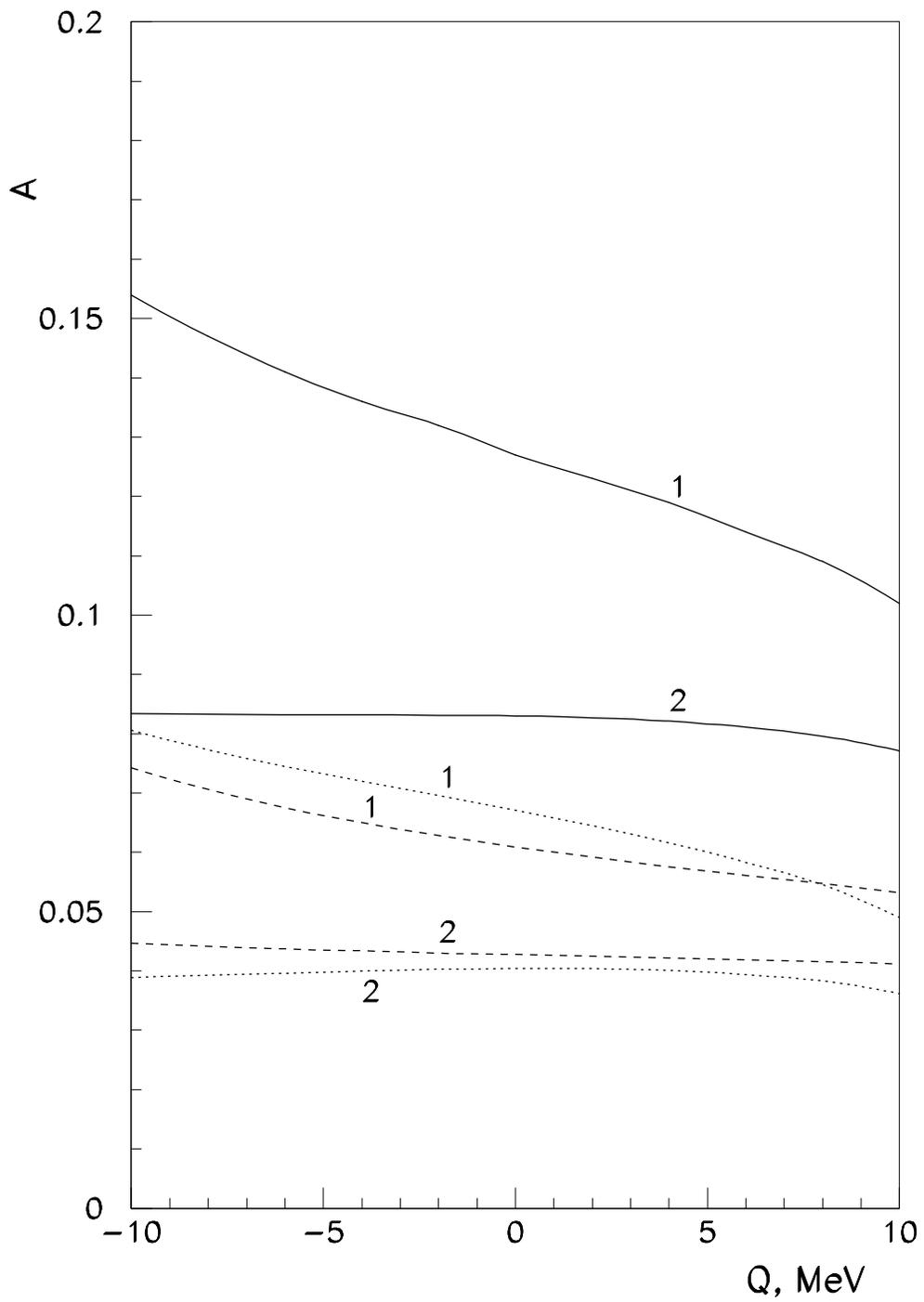
{\bf Fig.~5.} Plots of the asymmetry $A$ of $a^0_0$-meson production
     in reaction~(\ref{1}) vs. the energy release $Q$~(\ref{3}).
     Curves 1 and 2 correspond to two lower limits (specified by
     $C$$=$$1$ and $C$$=$$2$ (see text)) on the interval of $a^0_0$
     masses. The dashed (dotted) curves are calculated by taking account
     of the \mix\ mixing through only the \pieta\ transition ($K\bar K$
     decay channel). The solid curves are obtained with allowance made
     for both mixing mechanisms.

\end{document}